\documentclass[superscriptaddress,aps,nofootinbib,preprint]{revtex4-1}

\usepackage{verbatim}
\usepackage[T1]{fontenc}
\usepackage[utf8]{inputenc}
\usepackage[american]{babel}
\usepackage{epsfig}
\usepackage{graphicx,subcaption,caption}
\usepackage{booktabs}
\usepackage{multirow}
\usepackage{dcolumn}
\usepackage{amsmath}
\usepackage{mathtools}
\usepackage{amsfonts}
\usepackage{amssymb}
\usepackage{ulem}
\usepackage{epstopdf}
\usepackage{bm}
\usepackage{siunitx}
\usepackage{braket}
\usepackage{enumitem}
\usepackage{soul}
\usepackage[table]{xcolor}
\usepackage{color}
\usepackage{transparent}
\usepackage{pifont}
\usepackage{enumitem}
\usepackage{CJKutf8}

\definecolor{navyblue}{rgb}{0.0, 0.0, 0.5}
\definecolor{royalblue}{rgb}{0.25, 0.41, 0.88}
\definecolor{cadmiumgreen}{rgb}{0.0, 0.42, 0.24}
\definecolor{blue-violet}{rgb}{0.54, 0.17, 0.89}
\definecolor{darkviolet}{rgb}{0.58, 0.0, 0.83}
\definecolor{orange(colorwheel)}{rgb}{1.0, 0.5, 0.0}

\usepackage{hyperref}
\hypersetup{
    colorlinks=true, 
    linkcolor=royalblue, 
    citecolor=magenta}

\usepackage{booktabs}
\usepackage{multirow}
\usepackage{dcolumn}
\usepackage{colortbl}

\begin{document}

\title{Constraints on sterile neutrinos and the 
cosmological tensions}


\author{Supriya Pan}
\email{supriya.maths@presiuniv.ac.in}
\affiliation{Department of Mathematics, Presidency University, Kolkata 700073, India}
\affiliation{Institute of Systems Science, Durban University of Technology, PO Box 1334, Durban 4000, Republic of South Africa}

\author{Osamu Seto}
\email{seto@particle.sci.hokudai.ac.jp}
\affiliation{Department of Physics, Hokkaido University, Sapporo 060-0810, Japan}

\author{Tomo Takahashi}
\email{tomot@cc.saga-u.ac.jp}
\affiliation{Department of Physics, Saga University, Saga 840-8502, Japan}

\author{Yo Toda}
\email{y-toda@particle.sci.hokudai.ac.jp}
\affiliation{Department of Physics, Hokkaido University, Sapporo 060-0810, Japan}

\preprint{EPHOU-23-021} 

\begin{abstract}
We investigate cosmological bounds on sterile neutrino masses in the light of the Hubble and $S_8$ tensions. We argue that nonzero masses for sterile neutrinos are inferred at 2$\sigma$ level in some extended models such as  varying dark energy equation of state, when a direct measurement of the Hubble constant $H_0$ and weak lensing measurement of dark energy survey (DES) are taken into account. Furthermore, the Hubble and $S_8$ tensions are also reduced in such a framework. 
We also consider the case where a nonflat Universe is allowed and show that a slightly open Universe may be favored in models with sterile neutrinos in the context of the cosmological tensions.

\end{abstract}

\maketitle

\section{Introduction}

The large scale properties of our Universe are well described by 
the standard $\Lambda$-cold dark matter ($\Lambda$CDM) cosmological model, where the cosmological constant $\Lambda$ plays a role of dark energy whose equation of state (EoS) parameter $w$ corresponds to $-1$.
Although the $\Lambda$CDM model has been extremely successful in explaining the large scale structure of the Universe, however, there are several tensions between different measurements. One of the most serious issues which has been discussed much recently is the discrepancy of the Hubble constant $H_0$  between indirect and local direct measurements. The Planck measurement of the cosmic microwave background (CMB), the prime indirect measurements of $H_0$, infers $H_0 = (67.4 \pm 0.5)~{\rm km}/{\rm sec}/{\rm Mpc}$ at 68\% C.L.~\cite{Planck:2018vyg}, 
on the other hand, local direct measurements of $H_0$, such as those made using a distance ladder and strong gravitational lensing observations, consistently yield a value higher than that obtained from CMB~(e.g.,  \cite{Riess:2021jrx,Freedman:2020dne,Huang:2019yhh,Pesce:2020xfe,Wong:2019kwg}). 
For example, the Cepheid-supernovae distance ladder provides $H_0 = 73.04 \pm 1.04~{\rm km}/{\rm sec}/{\rm Mpc}$~\cite{Riess:2021jrx}.
For reviews on the Hubble constant tension, see e.g., Refs.~\cite{DiValentino:2021izs,Perivolaropoulos:2021jda,Schoneberg:2021qvd,Abdalla:2022yfr,Kamionkowski:2022pkx,Verde:2023lmm}.
Another tension is in the value of the amplitude of 
matter fluctuations, $\sigma_8$, at the scale of $8~{\rm  Mpc}/h$,  which is sometime recast as the $S_8$ parameter defined by $S_8 = \sigma_8\sqrt{\Omega_m/0.3}$. The  inferred value of $S_8$ from CMB by Planck~\cite{Planck:2018vyg} is larger than those measured by low red-shift probes through weak gravitational lensing and galaxy clustering~\cite{Joudaki:2019pmv,
DES:2021bvc,DES:2021vln,DES:2021wwk,KiDS:2021opn,Miyatake:2023njf,Sugiyama:2023fzm,Sunayama:2023hfm}.

Motivated by the issues mentioned above, it is worth investigating cosmological models beyond $\Lambda$CDM.
Although some simple extensions have been studied in the Planck 2018 paper~\cite{Planck:2018vyg}, such simple extended models cannot address the discrepancies. 
In particular, even when we only consider the data from Planck, they cannot simultaneously increase $H_0$ and decrease $\sigma_8$, which can be summarized as follows.
As the sum of neutrino mass $\sum m_{\nu}$ increases, both $H_0$ and $\sigma_8$ decrease.
As the curvature density parameter $\Omega_K$ decreases, both $H_0$ and $\sigma_8$ decrease for CMB data only, while it becomes consistent with a flat Universe when we include baryon acoustic oscillation (BAO) data.
As the effective number of neutrino species $N_\mathrm{eff}$ increases, both $H_0$ and $\sigma_8$ increase although eventually becomes
consistent with the prediction of the standard model of particle physics and  the Large Electron Positron Collider (LEP) results indicating  three generations of neutrinos~\cite{ALEPH:2005ab}. 
It should also be noted that none of these models cannot explain even a single tension when the combination of CMB and BAO data is adopted in the analysis.

Although three generations of neutrinos with nonvanishing masses and mixings explain various neutrino oscillation phenomena well in general, by examining each experiment in detail, it appears that there are multiple tensions between them. 
In addition,  the best-fit value of the mass squared difference for IceCube data is given as $\Delta m^2 \sim 4$ eV${}^2$~\cite{IceCube:2020phf} even though the data is consistent with no sterile neutrino. Furthermore, measurements of $\Delta m_{12}^2$ by the Super Kamiokande~\cite{Super-Kamiokande:2016yck}  and the KamLAND~\cite{KamLAND:2013rgu} also show mild discrepancy between them. Consequently, sterile neutrinos remain phenomenologically motivated  (see for instance \cite{Gariazzo:2013gua,Archidiacono:2014apa,Gariazzo:2014dla,RoyChoudhury:2018bsd,Hagstotz:2020ukm,Archidiacono:2020yey,DiValentino:2021rjj,Sakr:2022ans}),  despite that sterile neutrinos are challenged in other neutrino experiments~\cite{Maltoni:2002xd,DayaBay:2016lkk,Berryman:2021yan} and cosmological data~\cite{Hamann:2011ge,Planck:2018vyg}.

Cosmological constraints on such an eV mass sterile neutrino mostly come from CMB, big bang nucleosynthesis (BBN), and structure formation through the increase of $N_\mathrm{eff}$ and its free-streaming effect. Since right after the LSND results were published, incompatibility of such sterile neutrinos
 with BBN due to thermalization of sterile neutrino and the resultant large $N_\mathrm{eff}$ 
 has been pointed out~\cite{DiBari:2001ua,Barger:2003zg,Dolgov:2003sg}. 
However, by introducing large lepton asymmetry of $\mathcal{O}(10^{-2})$, which is permissible by current cosmological observations~\cite{Seto:2021tad}, one can relax the BBN constraint~\cite{Hamann:2011ge,Saviano:2013ktj}. 
A general equation of state (EoS) of dark energy~\cite{Kristiansen:2011mp} or an $f(R)$ theory of gravity~\cite{Motohashi:2010sj} could also eliminate the constraint on free streaming, which may still allow the sterile neutrinos.

As is well known, just introducing sterile neutrinos is not favored by the cosmological observations and it makes the fitting worse. 
Thus, we introduce an additional ingredient to  sterile neutrinos 
which might counteract or dominate over the effects from sterile neutrinos. 
As such an additional constituent, we consider the spatial curvature of the Universe,  which may be somewhat motivated
as it has been argued that CMB lensing magnitude indicates nonzero positive spatial curvature  at several standard deviations~\cite{Planck:2018vyg,DiValentino:2019qzk,Handley:2019tkm}. Another motivation might be that, in order that the varying electron mass can work as a solution to the Hubble tension \cite{Sekiguchi:2020teg}, which is  regarded as so far one of the best models 
to resolve the  tension~\cite{Schoneberg:2021qvd}, a positive spatial curvature would be preferable. 
This model is still consistent with BBN although the BBN data slightly prefers no variation of electron mass \cite{Seto:2022xgx}. Furthermore, it is interesting to note that larger neutrino masses are allowed in this framework \cite{Sekiguchi:2020igz}. 

Another simple extension is to consider some time-varying dark energy EoS rather than assuming a cosmological constant as dark energy. Although a simple time variation in the dark energy EoS such as the Chevallier-Polarski-Linder (CPL) parametrization \cite{Chevallier:2000qy,Linder:2002et} would not work as a solution to the Hubble tension, another EoS model may relax it (e.g., \cite{Keeley:2019esp}). Varying EoS  (also assuming a constant EoS)  may give a better fit in models with sterile neutrinos~\cite{Hamann:2011ge}.

In this paper, the fit of cosmological models with sterile neutrinos is examined, incorporating either the spatial curvature or assuming the CPL parametrization for dark energy EoS.
We perform our analysis for two different datasets: one set consists of CMB, BAO, and 
type Ia supernovae (SNeIa) while the other set additionally includes the SH0ES  (Supernovae and $H_0$ for the EoS of dark energy)~\cite{Riess:2021jrx} (which is referred to as ``R21'' hereafter) and Dark Energy Survey (DES) data \cite{DES:2017myr},
which give inconsistent values for $H_0$ and $\sigma_8 (S_8$), respectively. 
When we include R21 and DES, we show, by introducing additional parameters of the spatial curvature or the dark energy, a cosmological model  with sterile neutrinos appears to indicate larger $H_0$ and smaller $\sigma_8~(S_8)$ simultaneously.
We also find that a slightly negative curvature (open) Universe is favored 
when we consider a nonflat Universe
and eV-scale sterile neutrino mass is favored for a model with the CPL parametrization for the dark energy EoS.

This paper is organized as follows. 
In the next section, we summarize the method of our statistical analysis and the model framework beyond $\Lambda$CDM that we consider in this paper. In  Section.~\ref{sec:results}, we present our results. Section~\ref{sec:summary} is devoted to conclusion.

\section{Models, observational data and methodology}
\label{sec:model+data}

In this section, first we describe cosmological models to be analyzed to evaluate a possible contribution of sterile neutrinos. Then we give the methodology of our analysis and the data adopted in this paper to derive constraints on cosmological parameters. Bounds on cosmological parameters will be presented in the next section.

\subsection{Models to be analyzed}
\label{models}

\begin{enumerate}

\item  $\Lambda$CDM:  The base cosmological model which we use as the reference model to compare with other ones is the standard $\Lambda$CDM model with just six free parameters:
the baryon density $\omega_b=\Omega_b h^2$, the CDM density $\omega_c=\Omega_b h^2$, the amplitude and its spectral index of the primordial density perturbation, namely, $A_s$ and $n_s$ respectively, the acoustic scale  $\theta_* = r_* / D_M$ with $r_*$ the sound horizon at recombination and $D_M$  the angular diameter distance to the last scattering surface, and the optical depth of the reionization $\tau$.

\item $\Lambda$CDM$+$sterile neutrinos (labeled as ``Sterile''): To implement the contribution of sterile neutrino, we include and vary the effective number of neutrinos $N_{\mathrm{eff}}$ and the effective mass of the sterile neutrino $m_{\nu,\mathrm{sterile}}^{\mathrm{eff}} = (\Delta N_{\mathrm{eff}})^{3/4} m_{\nu,\mathrm{sterile}}^{\mathrm{thermal}} $ (in eV unit), as in \cite{Planck:2013pxb}. Note that the temperature fraction is given by $(T_S/T_\nu)^3= (\Delta N_{\mathrm{eff}})^{3/4} $ and the density parameter of sterile neutrino is given by $\omega_{\nu,\mathrm{sterile}}=m_{\nu,\mathrm{sterile}}^{\mathrm{eff}}/(94.1 \mathrm{eV})$.
Here, $\Delta N_{\mathrm{eff}} = N_{\mathrm{eff}}-N_{\mathrm{SM}}$ with the effective number of neutrinos in the standard model (SM) $N_{\mathrm{SM}}=3.046$, and $m_{\nu,\mathrm{sterile}}^{\mathrm{thermal}}$ is the physical mass for thermally produced sterile neutrinos.

\item $\Omega_K\Lambda$CDM$+$sterile neutrinos (labeled as  ``$\Omega_K$Sterile''): We consider an extension of $\Lambda$CDM$+$sterile neutrinos model by allowing the curvature of the Universe to
be varied by adding the curvature parameter $\omega_K=\Omega_K h^2$ in the analysis.

\item $w$CDM$+$sterile neutrinos (labeled as  ``$w$Sterile''):
Another extension of $\Lambda$CDM$+$sterile neutrinos model we consider is $w$CDM model in which the EoS of the dark energy $w$ is considered to be a free-to-vary parameter in an interval
with $w$ being constant.

\item $w w_a$CDM$+$sterile neutrinos (labeled as ``$w w_a$Sterile''): 
Yet another extension of $\Lambda$CDM$+$sterile neutrinos model is to assume a time-dependent EoS (the CPL parametrization \cite{Chevallier:2000qy,Linder:2002et}) which is written as 
\begin{align}
w(a)=w+w_a \left(1-\frac{a}{a_0}\right)=w+\frac{z}{1+z}w_a ,
\end{align}
where $a_0$ is the present value of the scale factor $a$, and 
$w$ and $w_a = dw/da$, are respectively the dark energy EoS at the present time and its derivative with respect to the scale factor.

\end{enumerate}

\subsection{datasets and methodology}

The observational datasets employed in this work are the following:

\begin{enumerate}

\item CMB data from Planck 2018~\cite{Planck:2019nip} (i.e. Planck TTTEEE+lowE) with 
CMB lensing~\cite{Planck:2018lbu}. 

\item  BAO distance measurements from various galaxy surveys~\cite{Beutler:2011hx, Ross:2014qpa, BOSS:2016wmc}. 

\item Local measurements of light curves and luminosity distance of type Ia supernovae, namely, the Pantheon sample~\cite{Pan-STARRS1:2017jku}. 

\item A prior on the Hubble constant 
($H_0 = 73.30 \pm 1.04 $ km/sec/Mpc at 68\% CL) from the SH0ES collaboration~\cite{Riess:2021jrx}. 

\item  Weak lensing measurement from the DES Year 1 data ~\cite{DES:2017myr,DES}.  We use the data from cosmic shear and galaxy clustering + galaxy-galaxy lensing. 

\end{enumerate}

In all analysis, we include the data from CMB (Planck), BAO and SNeIa (Pantheon), which we denote ``default" dataset in our analysis.
To constrain the parameters in the cosmological scenarios described above, we perform a Markov Chain Monte Carlo (MCMC) analysis by using $\texttt{\ensuremath{\mathrm{CosmoMC}}}$~\cite{Lewis:2002ah}
which is equipped with the Gelman-Rubin convergence diagnostic~\cite{Gelman:1992zz}.
We assume flat priors on all the parameters used in the MCMC analyses as shown in  Table~\ref{tab:priors}.  
In particular, we note that the sterile neutrino mass is restricted to 
$m_{\nu,\mathrm{sterile}}^{\mathrm{thermal}} < 10$ eV, following the analysis done in Planck 2018 \cite{Planck:2018vyg}.

\begin{table}
\begin{center}
\begin{tabular}{lccccccccc}
\hline
Parameter              &
$\Omega_{\rm b} h^2$   &
$\Omega_{\rm c} h^2$   & 
$\tau$      & 
$n_s$                  & 
$\ln{(10^{10} A_{s})}$ & 
$\theta_{\ast}$        & 
\\
 Prior   & 
 $[0.005, 0.1]$ & 
 $[0.001, 0.99]$ & 
 $[0.01, 0.8]$   &
 $[0.8, 1.2]$    &
 $[1.61, 3.91]$  &
 $[0.5, 10]$     &
 \\
\hline
\end{tabular}
 \\

\begin{tabular}{lccccccccc}
\hline
Parameter              &
$N_{\rm eff}$          & 
$m_{\nu,{\rm {sterile}}}^{{\rm {eff}}}$ & 
$\Omega_K$             &
$w$                    &
$w_a$           \\
 Prior   & 
 $[3.0461, 4]$   &
 $[0, 10]$       &
 $[-0.3, 0.3]$  &
 $[-2, 0]$  &
 $[-2, 1]$
  \\
\hline
\end{tabular}
\end{center}
\caption{Cosmological parameters and the prior range adopted in the analysis. We assume a flat prior for all the parameters.}
\label{tab:priors}
\end{table}

\maketitle
\begin{table}
\begin{tabular}{lcccc}
\hline 
 &  & \multicolumn{2}{c}{$\Lambda\mathrm{CDM}$} & \tabularnewline
\hline 
Parameter  & $\mathcal{D}$+Pantheon & $\mathcal{D}$+R21 & $\mathcal{D}$+Pantheon+DES  & $\mathcal{D}$+R21+DES\tabularnewline
\hline 
{\boldmath\ensuremath{\Omega_{b}h^{2}}}  & \ensuremath{0.02244\pm0.00013}  & $\ensuremath{0.02259\pm0.00013}$ & $\ensuremath{0.02252\pm0.00013}$ & $\ensuremath{0.02265\pm0.00013}$\tabularnewline
{\boldmath\ensuremath{\Omega_{c}h^{2}}}  & \ensuremath{0.11918\pm0.00094}  & $\ensuremath{0.11765\pm0.00088}$ & $\ensuremath{0.11807\pm0.00084}$ & $0.11669\pm0.00082$\tabularnewline
{\boldmath\ensuremath{100\theta_{MC}}}  & \ensuremath{1.04103\pm0.00028} & $\ensuremath{1.04126\pm0.00029}$ & $\ensuremath{1.04111\pm0.00029}$ & $1.04132\pm0.00029$\tabularnewline
{\boldmath\ensuremath{\tau}}  & \ensuremath{0.0562_{-0.0079}^{+0.0068}}  & $\ensuremath{0.0613_{-0.0082}^{+0.0069}}$ & $\ensuremath{0.0572_{-0.0084}^{+0.0068}}$ & $\ensuremath{0.0580_{-0.0083}^{+0.0071}}$\tabularnewline
{\boldmath\ensuremath{{\rm {ln}}(10^{10}A_{s})}}  & \ensuremath{3.047_{-0.016}^{+0.013}}  & $\ensuremath{3.054_{-0.016}^{+0.014}}$ & \ensuremath{3.046_{-0.017}^{+0.013}} & $\ensuremath{3.044_{-0.017}^{+0.015}}$\tabularnewline
{\boldmath\ensuremath{n_{s}}}  & \ensuremath{0.9670\pm0.0038}  & $\ensuremath{0.9710\pm0.0036}$ & $\ensuremath{0.9692\pm0.0036}$ & $\ensuremath{0.9729\pm0.0036}$\tabularnewline
\ensuremath{H_{0}}  & \ensuremath{67.74\pm0.42}  & $\ensuremath{68.49\pm0.40}$ & $\ensuremath{68.23\pm0.38}$ & $\ensuremath{68.90\pm0.38}$\tabularnewline
\ensuremath{S_{8}}  & \ensuremath{0.8234_{-0.011}^{+0.0098}}  & $\ensuremath{0.8090\pm0.0099}$ & $\ensuremath{0.8108\pm0.0091}$ & $\ensuremath{0.795\pm0.010}$\tabularnewline
\hline 
\end{tabular}

\begin{tabular}{lcccc}
\hline 
 &  & \multicolumn{2}{c}{Sterile} & \tabularnewline
\hline 
Parameter  & $\mathcal{D}$+Pantheon  & $\mathcal{D}$+R21 & $\mathcal{D}$+Pantheon+DES  & $\mathcal{D}$+R21+DES\tabularnewline
\hline 
{\boldmath\ensuremath{\Omega_{b}h^{2}}}  & \ensuremath{0.02249\pm0.00014} & $\ensuremath{0.02282\pm0.00016}$ & $\ensuremath{0.02256\pm0.00014}$ & $\ensuremath{0.02286\pm0.00015}$\tabularnewline
{\boldmath\ensuremath{\Omega_{c}h^{2}}}  & \ensuremath{0.1189_{-0.0022}^{+0.0033}} & $\ensuremath{0.1245_{-0.0025}^{+0.0031}}$ & $\ensuremath{0.1170_{-0.0020}^{+0.0025}}$ & $\ensuremath{0.1223_{-0.0022}^{+0.0030}}$\tabularnewline
{\boldmath\ensuremath{100\theta_{MC}}}  & \ensuremath{1.04091\pm0.00033} & $\ensuremath{1.04048\pm0.00041}$ & $\ensuremath{1.04105\pm0.00030}$ & $\ensuremath{1.04068\pm0.00040}$\tabularnewline
{\boldmath\ensuremath{\tau}}  & \ensuremath{0.0580\pm0.0071} & $\ensuremath{0.0615_{-0.0084}^{+0.0071}}$ & $\ensuremath{0.0588_{-0.0080}^{+0.0069}}$ & $\ensuremath{0.0588_{-0.0086}^{+0.0075}}$\tabularnewline
{\boldmath\ensuremath{{\rm {ln}}(10^{10}A_{s})}}  & \ensuremath{3.053\pm0.015} & $\ensuremath{3.071\pm0.016}$ & $\ensuremath{3.052_{-0.016}^{+0.014}}$ & $\ensuremath{3.059\pm0.017}$\tabularnewline
{\boldmath\ensuremath{n_{s}}}  & \ensuremath{0.9688_{-0.0054}^{+0.0041}} & $\ensuremath{0.9832_{-0.0052}^{+0.0060}}$ & $\ensuremath{0.9685_{-0.0047}^{+0.0039}}$ & $\ensuremath{0.9835_{-0.0052}^{+0.0062}}$\tabularnewline
{\boldmath\ensuremath{N_{\rm eff}}}  & \ensuremath{<3.13} & $\ensuremath{3.46_{-0.14}^{+0.16}}$ & $\ensuremath{3.118_{-0.059}^{+0.015}}$ & $\ensuremath{3.40_{-0.13}^{+0.16}}$\tabularnewline
{\boldmath\ensuremath{m_{\nu,{\rm {sterile}}}^{{\rm eff}}}}  & \ensuremath{<0.194} & $\ensuremath{<0.0379}$ & $\ensuremath{0.29_{-0.25}^{+0.10}}$ & $\ensuremath{<0.0760}$\tabularnewline
\ensuremath{H_{0}}  & \ensuremath{67.89_{-0.69}^{+0.40}}  & $\ensuremath{70.49_{-0.81}^{+0.94}}$ & $\ensuremath{67.98_{-0.51}^{+0.40}}$ & $\ensuremath{70.46_{-0.79}^{+0.98}}$\tabularnewline
\ensuremath{S_{8}}  & \ensuremath{0.812_{-0.014}^{+0.019}}  & $\ensuremath{0.820_{-0.011}^{+0.014}}$ & $\ensuremath{0.790\pm0.015}$ & $\ensuremath{0.798_{-0.012}^{+0.015}}$\tabularnewline
\hline 
\end{tabular}\caption{ Marginalized values and $68\%$ confidence regions for cosmological parameters obtained combining Planck and BAO  with and without
other datasets (e.g. Pantheon, R21, DES), for $\Lambda$CDM model and Sterile model. Here \textquotedblleft$\mathcal{D}$"
means the combined dataset with Planck and BAO.}
\label{tab:LCDM+Sterile} 
\end{table}
\begin{table}
\begin{tabular}{lcccc}
\hline 
 &  & \multicolumn{2}{c}{$\ensuremath{\Omega_{K}}$Sterile} & \tabularnewline
\hline 
Parameter  & $\mathcal{D}$+Pantheon  & $\mathcal{D}$+R21  & $\mathcal{D}$+Pantheon+DES  & $\mathcal{D}$+R21+DES\tabularnewline
\hline 
{\boldmath\ensuremath{\Omega_{b}h^{2}}}  & $\ensuremath{0.02247_{-0.00017}^{+0.00016}}$ & $\ensuremath{0.02261\pm0.00021}$ & $\ensuremath{0.02249\pm0.00016}$ & $\ensuremath{0.02256_{-0.00019}^{+0.00017}}$\tabularnewline
{\boldmath\ensuremath{\Omega_{c}h^{2}}}  & $\ensuremath{0.1195_{-0.0024}^{+0.0033}}$ & $\ensuremath{0.1238\pm0.0029}$ & $\ensuremath{0.1185_{-0.0022}^{+0.0027}}$ & $\ensuremath{0.1222\pm0.0028}$\tabularnewline
{\boldmath\ensuremath{100\theta_{MC}}}  & $\ensuremath{1.04082\pm0.00036}$ & $\ensuremath{1.04050\pm0.00039}$ & $\ensuremath{1.04087\pm0.00034}$ & $\ensuremath{1.04059\pm0.00039}$\tabularnewline
{\boldmath\ensuremath{\tau}}  & $\ensuremath{0.0575_{-0.0078}^{+0.0068}}$ & $\ensuremath{0.0603_{-0.0086}^{+0.0069}}$ & $\ensuremath{0.0585_{-0.0080}^{+0.0069}}$ & $\ensuremath{0.0576\pm0.0079}$\tabularnewline
{\boldmath\ensuremath{{\rm {ln}}(10^{10}A_{s})}}  & $\ensuremath{3.054\pm0.015}$ & $\ensuremath{3.067_{-0.018}^{+0.015}}$ & $\ensuremath{3.055\pm0.015}$ & $\ensuremath{3.058\pm0.017}$\tabularnewline
{\boldmath\ensuremath{n_{s}}}  & $\ensuremath{0.9671_{-0.0066}^{+0.0049}}$ & $\ensuremath{0.9744\pm0.0085}$ & $\ensuremath{0.9659_{-0.0057}^{+0.0049}}$ & $\ensuremath{0.9702_{-0.0088}^{+0.0062}}$\tabularnewline
{\boldmath\ensuremath{N_{\rm eff}}}  & $\ensuremath{<3.15}$ & $\ensuremath{3.33_{-0.20}^{+0.11}}$ & $\ensuremath{3.137_{-0.073}^{+0.025}}$ & $3.279_{-0.17}^{+0.092}$\tabularnewline
{\boldmath\ensuremath{m_{\nu,{\rm {sterile}}}^{{\rm {eff}}}}}  & $\ensuremath{<0.231}$ & $\ensuremath{<0.0855}$ & $\ensuremath{0.32_{-0.20}^{+0.14}}$ & $\ensuremath{0.28_{-0.20}^{+0.14}}$\tabularnewline
{\boldmath\ensuremath{\Omega_{K}}}  & $\ensuremath{0.0013\pm0.0021}$ & $\ensuremath{0.0034\pm0.0025}$ & $\ensuremath{0.0024_{-0.0023}^{+0.0021}}$ & $\ensuremath{0.0057_{-0.0022}^{+0.0026}}$\tabularnewline
\ensuremath{H_{0}}  & $\ensuremath{68.25_{-0.80}^{+0.68}}$ & $\ensuremath{70.40\pm0.80}$ & $\ensuremath{68.54\pm0.69}$ & $\ensuremath{70.41_{-0.76}^{+0.67}}$\tabularnewline
\ensuremath{S_{8}}  & $\ensuremath{0.809_{-0.014}^{+0.021}}$ & $0.811_{-0.011}^{+0.019}$ & $\ensuremath{0.787\pm0.015}$ & $\ensuremath{0.777\pm0.018}$\tabularnewline
\hline 
\end{tabular}

\caption{Marginalized values and $68\%$ confidence regions for cosmological
parameters obtained combining Planck and BAO with and without
other datasets (e.g. Pantheon, R21, DES), for $\Omega_{K}$Sterile model.}
\label{tab:Kwwa} 
\end{table}
\begin{figure}
\includegraphics[width=15cm]{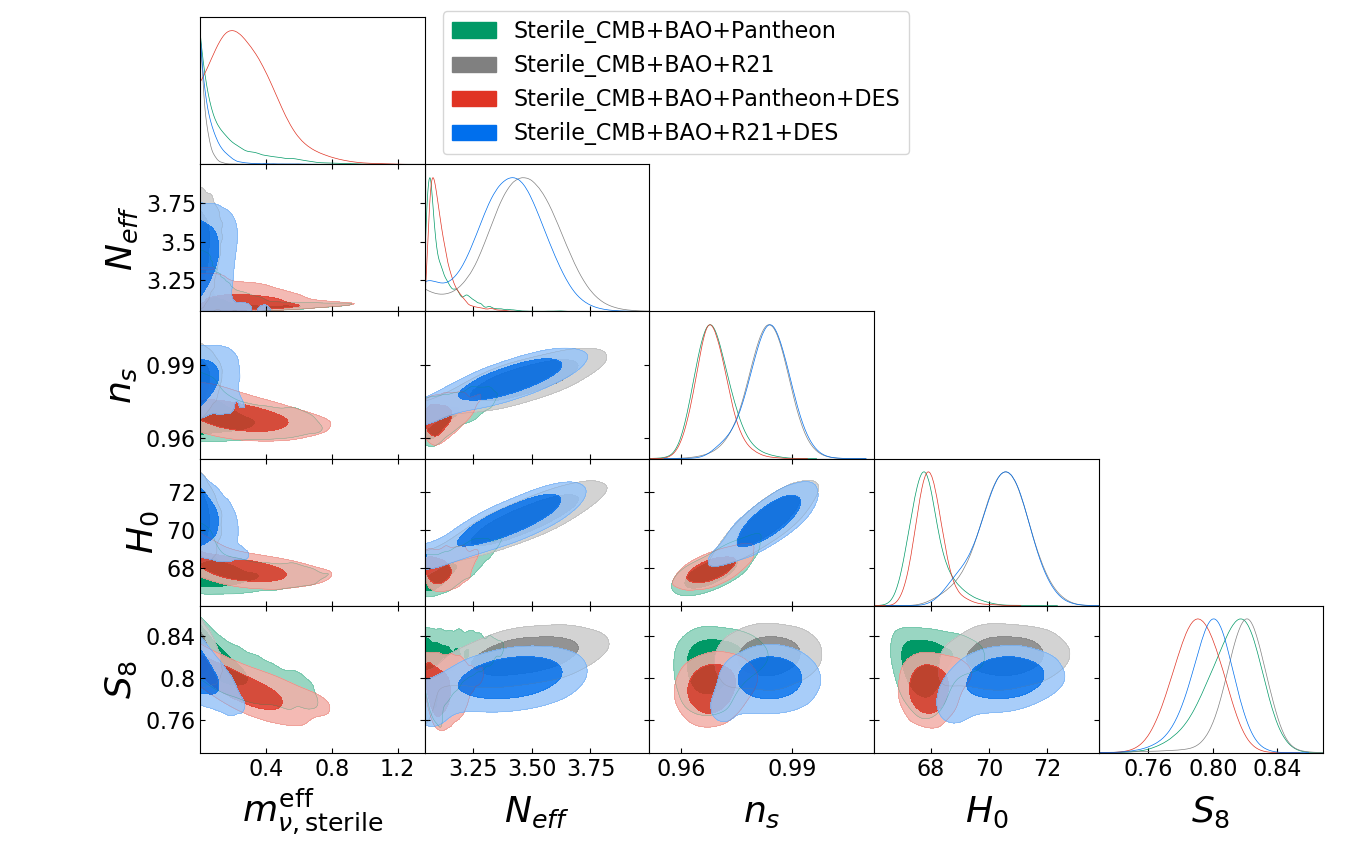}
\caption{
Posterior distributions of some parameters for the ``Sterile'' model. Different colors of greenish, grayish,  reddish and bluish contours stand for different datasets. 
}
\label{fig:sterile}
\end{figure}
\begin{figure}
\includegraphics[width=15cm]{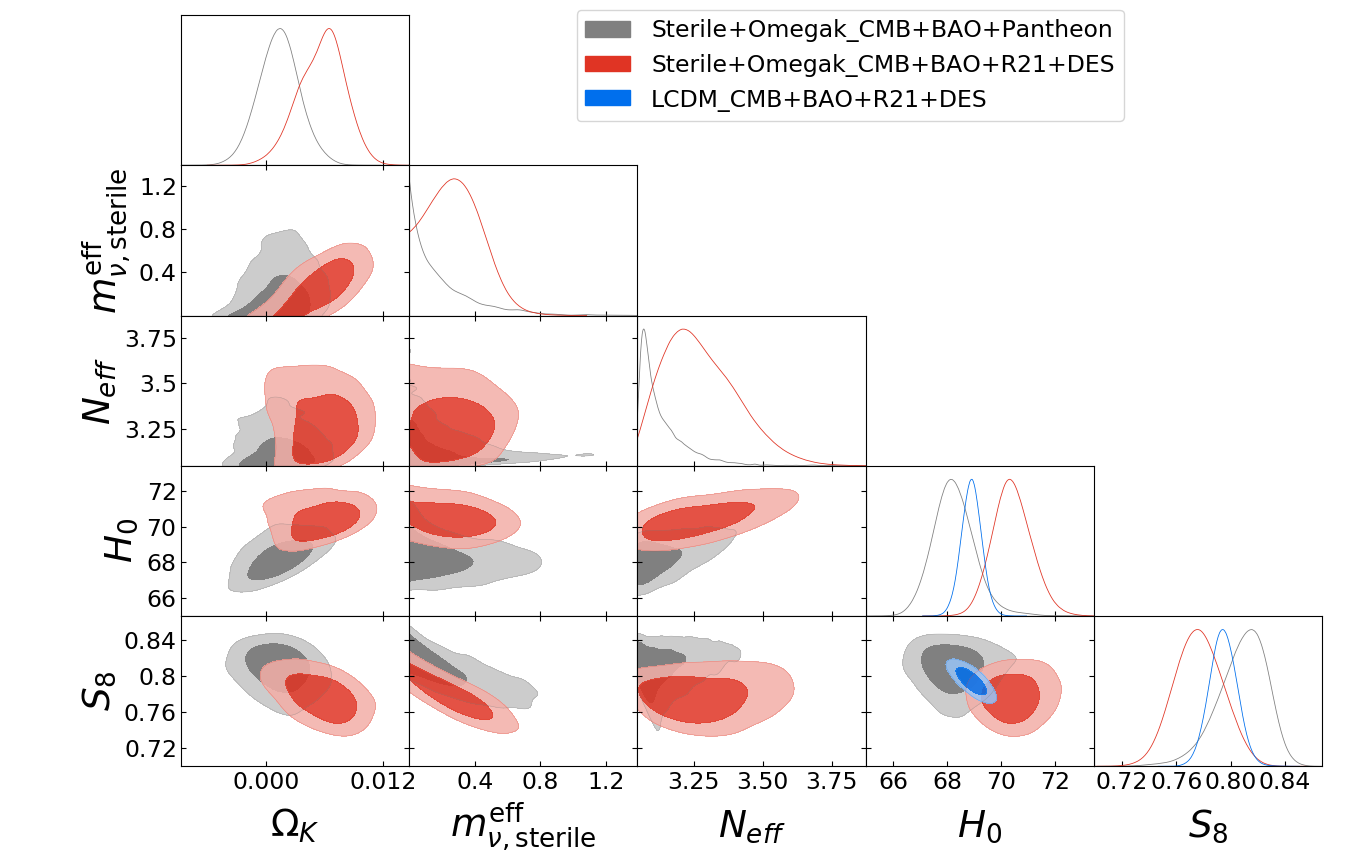}
\caption{
Posterior distributions of some parameters for ``$\Omega_K$Sterile'' model.
Different colors of grayish and reddish contours stand for different datasets. 
For comparison, that of $\Lambda$CDM model is also shown with bluish contour.
}
\label{fig:OmegaK}
\end{figure}
\begin{figure}
\includegraphics[width=15cm]{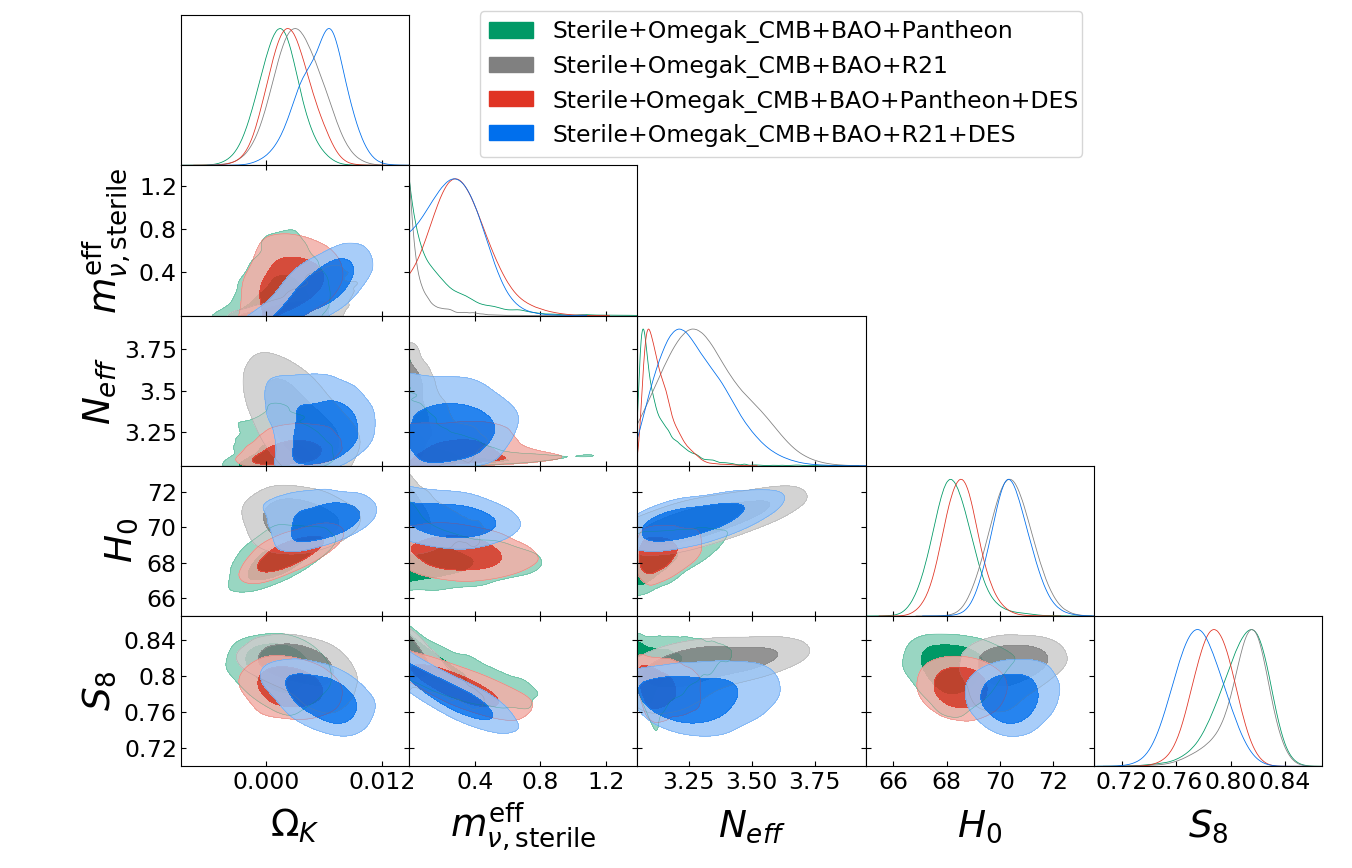}
\caption{
Posterior distributions of some parameters for ``$\Omega_K$Sterile'' model.
Different colors of greenish, grayish, reddish and bluish contours stand for different datasets. 
}
\label{fig:OmegaK2}
\end{figure}
%
\begin{figure}
\includegraphics[width=15cm]{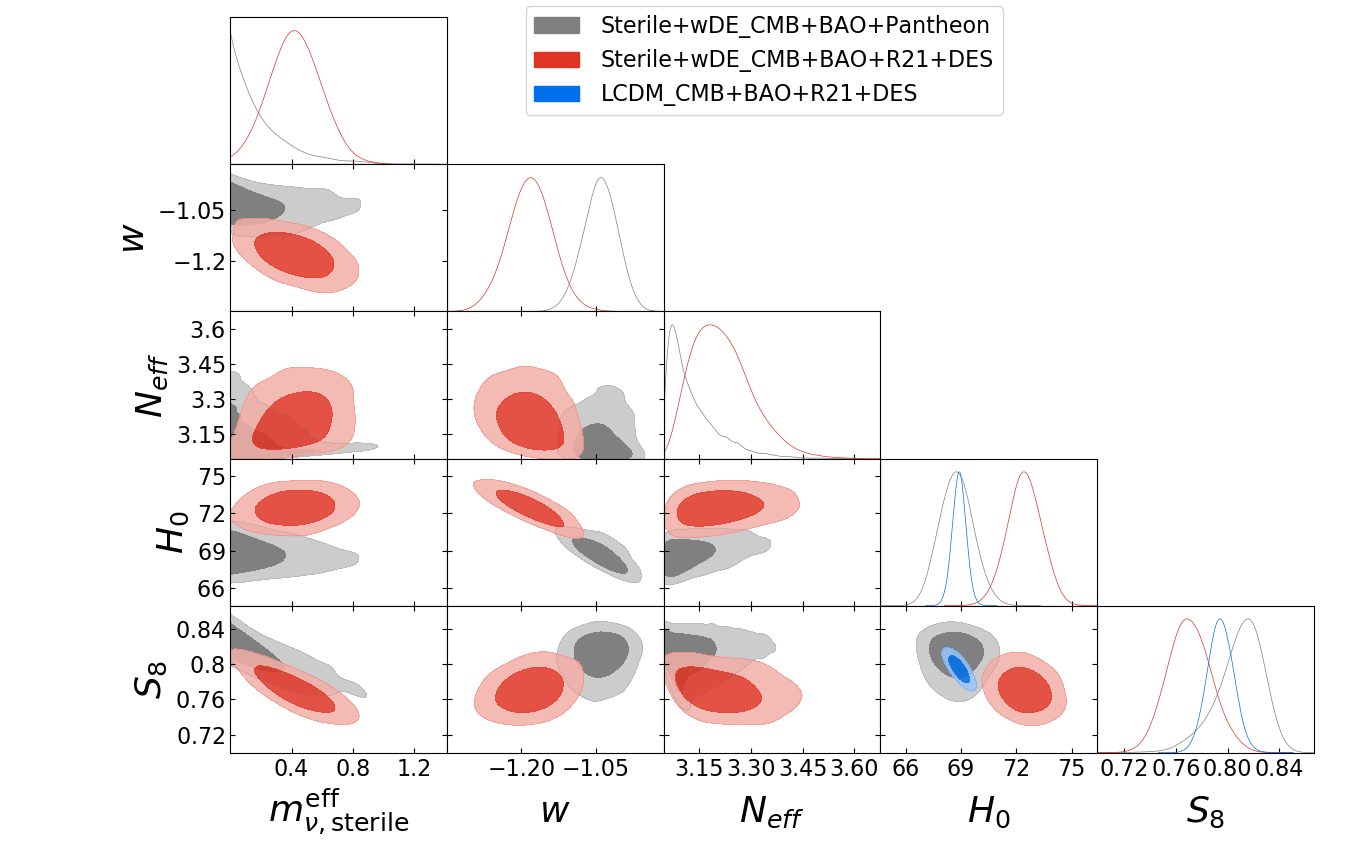}
\caption{Posterior distributions of some parameters for the ``$w$Sterile'' model. Different colors of grayish, reddish and bluish contours stand for different datasets. }
\label{fig:wsterile}
\end{figure}
\begin{figure}
\includegraphics[width=15cm]{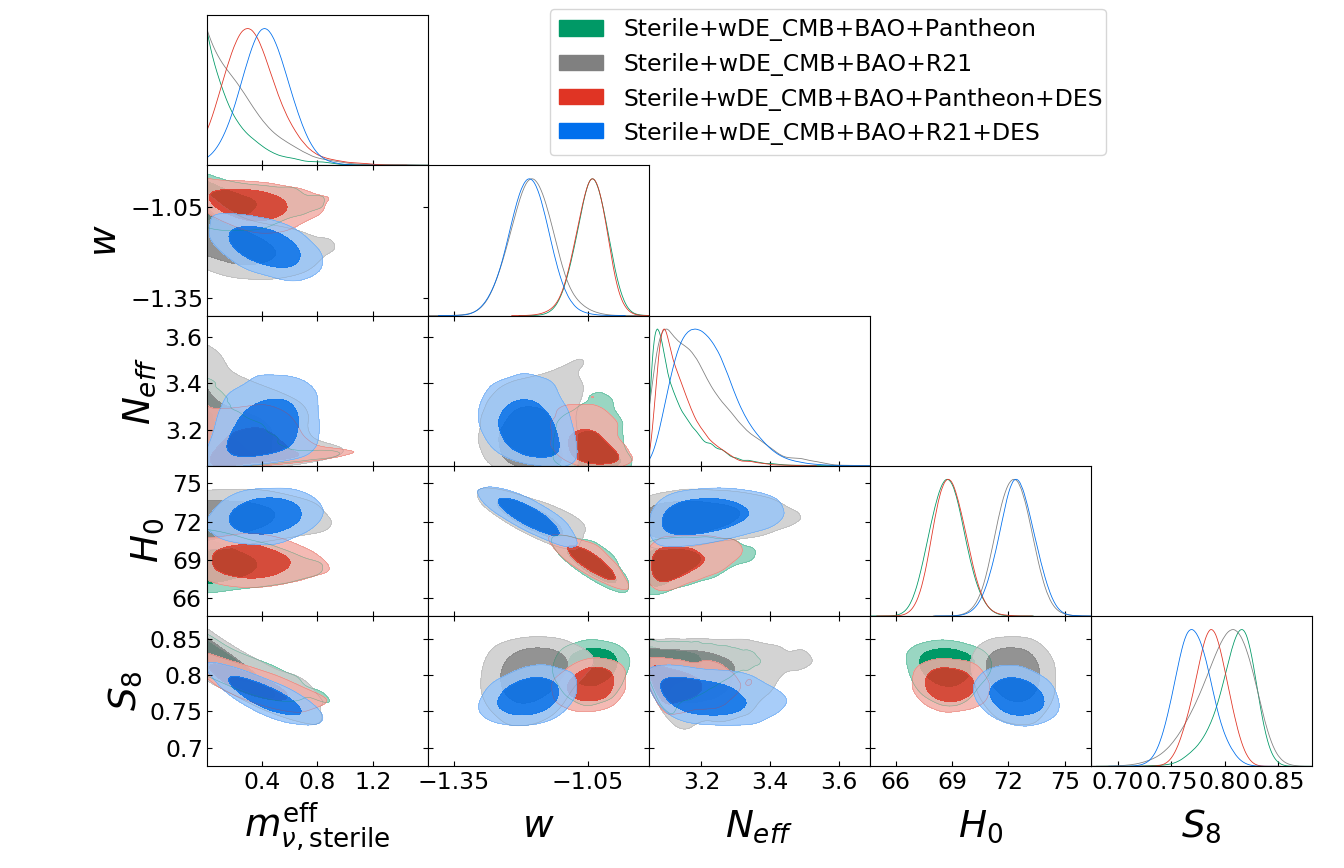}
\caption{Posterior distributions of some parameters for the ``$w$Sterile'' model. Different colors of greenish, grayish, reddish and bluish contours stand for different datasets. }
\label{fig:wsterile2}
\end{figure}
\begin{figure}
\includegraphics[width=15cm]{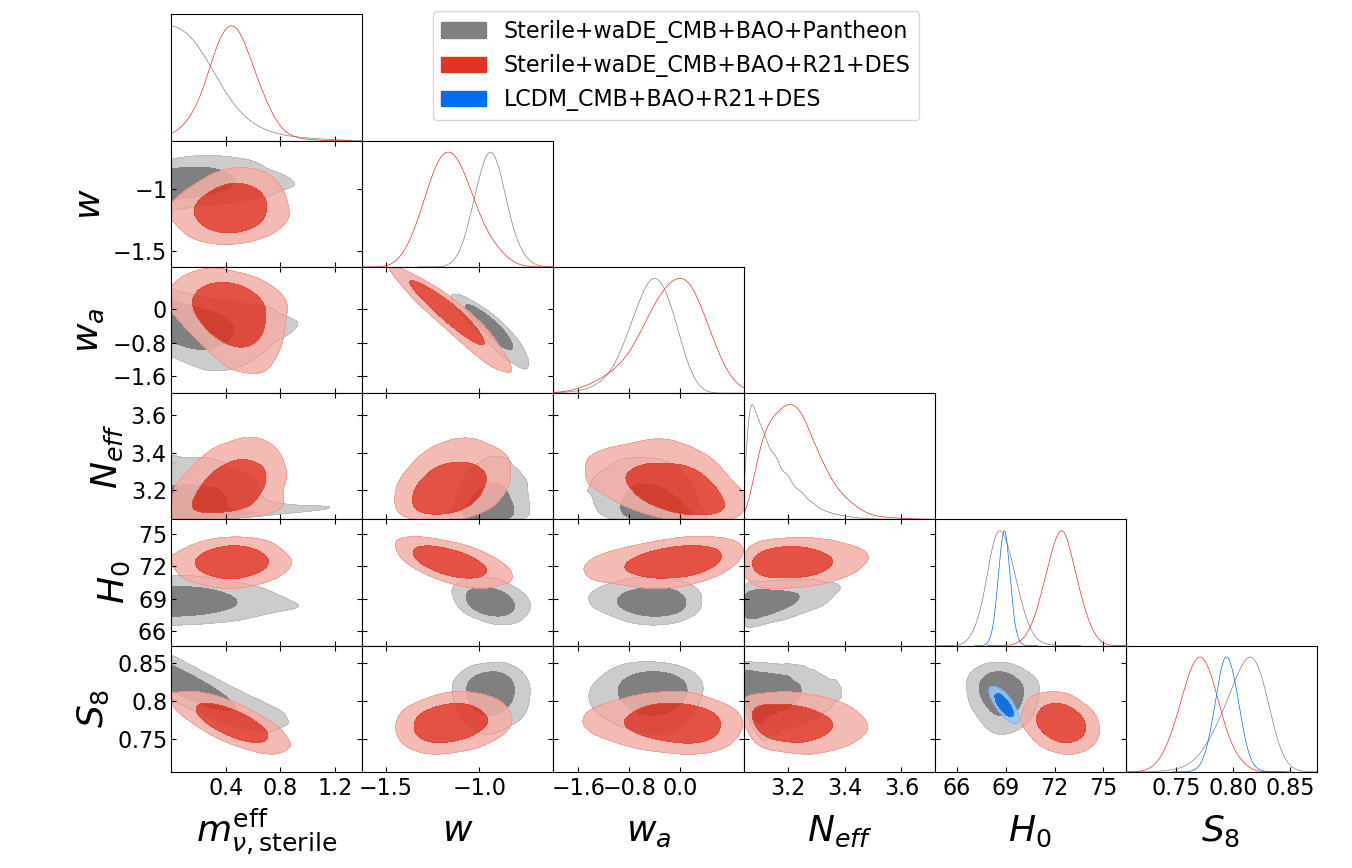}
\caption{Posterior distributions of some parameters for ``$ww_a$Sterile'' model.
Different colors of grayish and reddish contours stand for different datasets. 
For comparison, that of $\Lambda$CDM model is also shown with bluish contour.}
\label{fig:wwa}
\end{figure}
%
\begin{table}
\begin{tabular}{lcccc}
\hline 
 &  & \multicolumn{2}{c}{$\ensuremath{w}$Sterile} & \tabularnewline
\hline 
Parameter  & $\mathcal{D}$+Pantheon  & $\mathcal{D}$+R21 & $\mathcal{D}$+pantheon+DES  & $\mathcal{D}$+R21+DES \tabularnewline
\hline 
{\boldmath\ensuremath{\Omega_{b}h^{2}}}  & $\ensuremath{0.02246\pm0.00015}$ & $\ensuremath{0.02246_{-0.00017}^{+0.00015}}$ & $\ensuremath{0.02253\pm0.00015}$ & $\ensuremath{0.02255\pm0.00015}$\tabularnewline
{\boldmath\ensuremath{\Omega_{c}h^{2}}}  & $\ensuremath{0.1194_{-0.0018}^{+0.0031}}$ & $\ensuremath{0.1209_{-0.0023}^{+0.0032}}$ & $\ensuremath{0.1176_{-0.0019}^{+0.0028}}$ & $\ensuremath{0.1196\pm0.0022}$\tabularnewline
{\boldmath\ensuremath{100\theta_{MC}}}  & $\ensuremath{1.04083_{-0.00031}^{+0.00034}}$ & $\ensuremath{1.04068\pm0.00035}$ & $\ensuremath{1.04093\pm0.00032}$ & $\ensuremath{1.04079\pm0.00032}$\tabularnewline
{\boldmath\ensuremath{\tau}}  & $\ensuremath{0.0561\pm0.0078}$ & $\ensuremath{0.0549\pm0.0079}$ & $\ensuremath{0.0572_{-0.0080}^{+0.0070}}$ & $\ensuremath{0.0545\pm0.0080}$\tabularnewline
{\boldmath\ensuremath{{\rm {ln}}(10^{10}A_{s})}}  & $\ensuremath{3.050_{-0.016}^{+0.015}}$ & $\ensuremath{3.051\pm0.017}$ & $\ensuremath{3.051_{-0.016}^{+0.014}}$ & $\ensuremath{3.048\pm0.017}$\tabularnewline
{\boldmath\ensuremath{n_{s}}}  & $\ensuremath{0.9672_{-0.0055}^{+0.0044}}$ & $\ensuremath{0.9666_{-0.0067}^{+0.0048}}$ & $\ensuremath{0.9672_{-0.0050}^{+0.0043}}$ & $\ensuremath{0.9668_{-0.0058}^{+0.0047}}$\tabularnewline
{\boldmath\ensuremath{N_{\rm eff}}}  & $\ensuremath{3.132_{-0.085}^{+0.014}}$ & $\ensuremath{3.192_{-0.14}^{+0.039}}$ & $\ensuremath{3.140_{-0.076}^{+0.023}}$ & $\ensuremath{3.216_{-0.10}^{+0.061}}$\tabularnewline
{\boldmath\ensuremath{m_{\nu,{\rm {sterile}}}^{{\rm {eff}}}}}  & $\ensuremath{<0.236}$ & $\ensuremath{<0.326}$ & $\ensuremath{0.35_{-0.23}^{+0.14}}$ & $\ensuremath{0.43\pm0.17}$\tabularnewline
{\boldmath\ensuremath{w}}  & $\ensuremath{-1.041\pm0.035}$ & $\ensuremath{-1.175\pm0.049}$ & $\ensuremath{-1.044_{-0.032}^{+0.037}}$ & $\ensuremath{-1.182\pm0.045}$\tabularnewline
\ensuremath{H_{0}}  & $\ensuremath{68.75\pm0.92}$ & $\ensuremath{72.29\pm0.93}$ & $\ensuremath{68.89_{-0.92}^{+0.82}}$ & $\ensuremath{72.44\pm0.93}$\tabularnewline
\ensuremath{S_{8}}  & $\ensuremath{0.810_{-0.013}^{+0.021}}$ & $\ensuremath{0.800_{-0.020}^{+0.030}}$ & $\ensuremath{0.787\pm0.015}$ & $\ensuremath{0.771\pm0.017}$\tabularnewline
\hline 
\end{tabular}

\begin{tabular}{lcccc}
\hline 
 &  & \multicolumn{2}{c}{$\ensuremath{ww_{a}}$Sterile} & \tabularnewline
\hline 
Parameter  & $\mathcal{D}$+Pantheon  & $\mathcal{D}$+R21 & $\mathcal{D}$+Pantheon+DES  & $\mathcal{D}$+R21+DES \tabularnewline
\hline 
{\boldmath\ensuremath{\Omega_{b}h^{2}}}  & $\ensuremath{0.02244\pm0.00015}$ & $\ensuremath{0.02246\pm0.00016}$ & $\ensuremath{0.02252\pm0.00015}$ & $\ensuremath{0.02254\pm0.00015}$\tabularnewline
{\boldmath\ensuremath{\Omega_{c}h^{2}}}  & $\ensuremath{\ensuremath{0.1199_{-0.0020}^{+0.0032}}}$ & $\ensuremath{0.1210_{-0.0024}^{+0.0031}}$ & $\ensuremath{0.1189_{-0.0021}^{+0.0028}}$ & $\ensuremath{0.1198\pm0.0025}$\tabularnewline
{\boldmath\ensuremath{100\theta_{MC}}}  & $1.04076_{-0.00031}^{+0.00035}$ & $\ensuremath{1.04068_{-0.00033}^{+0.00036}}$ & $\ensuremath{1.04081\pm0.00033}$ & $\ensuremath{1.04076\pm0.00033}$\tabularnewline
{\boldmath\ensuremath{\tau}}  & $\ensuremath{0.0542\pm0.0074}$ & $\ensuremath{0.0546_{-0.0082}^{+0.0074}}$ & $\ensuremath{0.0555\pm0.0076}$ & $\ensuremath{0.0545\pm0.0079}$\tabularnewline
{\boldmath\ensuremath{{\rm {ln}}(10^{10}A_{s})}}  & $\ensuremath{3.047\pm0.015}$ & $\ensuremath{3.050_{-0.018}^{+0.016}}$ & $\ensuremath{3.049\pm0.016}$ & $\ensuremath{3.049\pm0.017}$\tabularnewline
{\boldmath\ensuremath{n_{s}}}  & $\ensuremath{0.9663_{-0.0055}^{+0.0044}}$ & $\ensuremath{0.9663_{-0.0067}^{+0.0050}}$ & $\ensuremath{0.9663_{-0.0052}^{+0.0044}}$ & $\ensuremath{0.9666_{-0.0059}^{+0.0047}}$\tabularnewline
{\boldmath\ensuremath{N_{\rm eff}}}  & $\ensuremath{\ensuremath{3.151_{-0.10}^{+0.025}}}$ & $3.190_{-0.14}^{+0.039}$ & $\ensuremath{3.183_{-0.11}^{+0.043}}$ & $\ensuremath{3.229_{-0.12}^{+0.065}}$\tabularnewline
{\boldmath\ensuremath{m_{\nu,{\rm {sterile}}}^{{\rm {eff}}}}}  & $\ensuremath{\ensuremath{<0.318}}$ & $\ensuremath{<0.330}$ & $\ensuremath{0.40_{-0.20}^{+0.16}}$ & $\ensuremath{0.45\pm0.18}$\tabularnewline
{\boldmath\ensuremath{w}}  & $\ensuremath{-0.939\pm0.085}$ & $\ensuremath{-1.19_{-0.15}^{+0.12}}$ & $\ensuremath{-0.925_{-0.090}^{+0.079}}$ & $\ensuremath{-1.15_{-0.14}^{+0.12}}$\tabularnewline
{\boldmath\ensuremath{w_{a}}}  & $\ensuremath{\ensuremath{-0.46_{-0.31}^{+0.40}}}$ & $\ensuremath{0.03_{-0.38}^{+0.57}}$ & $\ensuremath{-0.56_{-0.31}^{+0.43}}$ & $\ensuremath{-0.15_{-0.41}^{+0.58}}$\tabularnewline
\ensuremath{H_{0}}  & $\ensuremath{\ensuremath{68.76_{-0.93}^{+0.84}}}$ & $\ensuremath{72.35\pm0.99}$ & $\ensuremath{69.02\pm0.89}$ & $\ensuremath{72.39\pm0.98}$\tabularnewline
\ensuremath{S_{8}}  & $\ensuremath{\ensuremath{0.809_{-0.016}^{+0.022}}}$ & $\ensuremath{0.799_{-0.021}^{+0.028}}$ & $\ensuremath{0.787\pm0.015}$ & $\ensuremath{0.771\pm0.017}$\tabularnewline
\hline 
\end{tabular}
\caption{Marginalized values and $68\%$ confidence regions for the cosmological
parameters obtained combining Planck and BAO with and without
other datasets (e.g. Pantheon, R21, DES), for \textquotedblleft$w$Sterile\textquotedblright{}
and \textquotedblleft$ww_{a}$Sterile\textquotedblright{} models. }
\label{tab:wwa} 
\end{table}

\begin{figure}
\includegraphics[width=15cm]{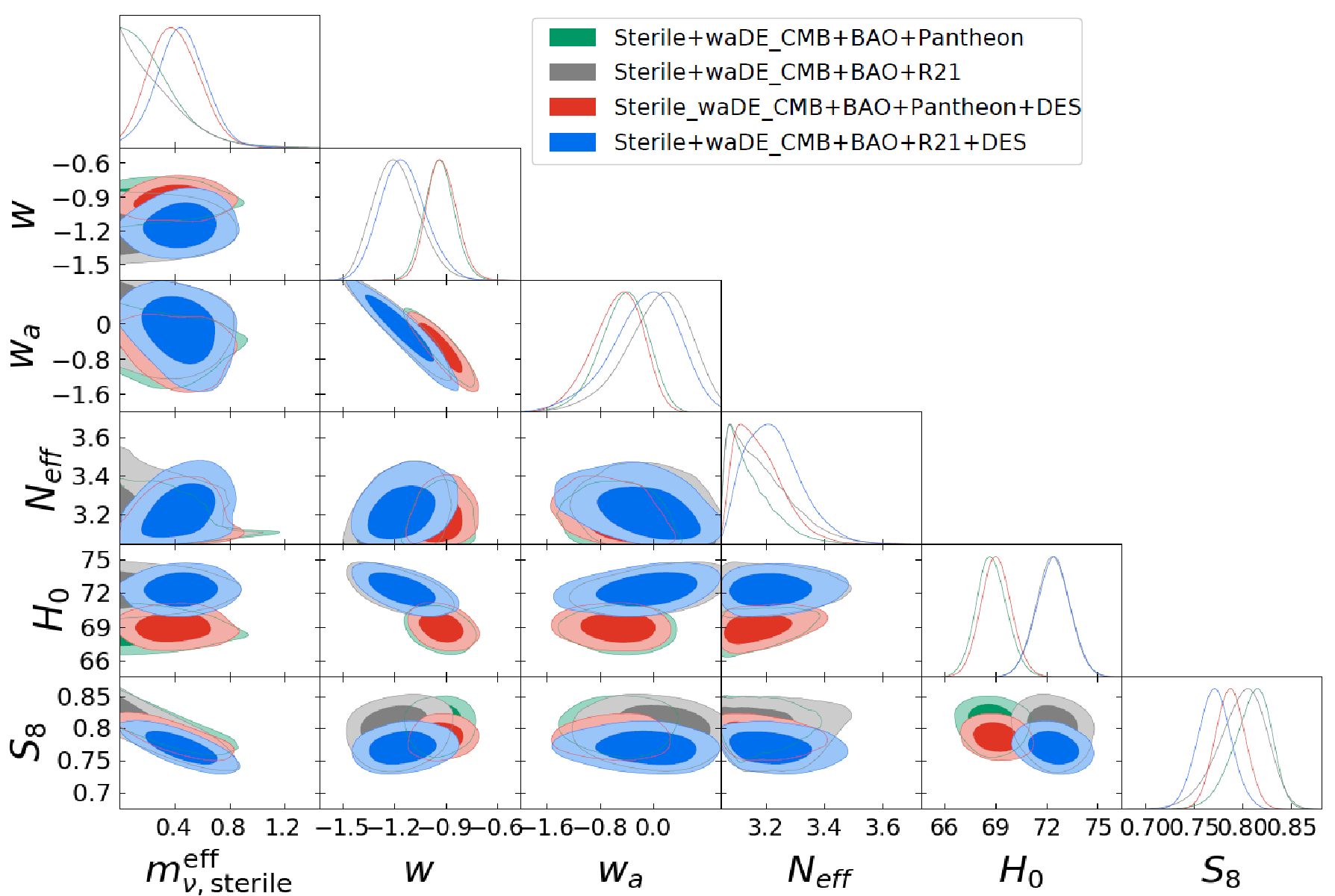}
\caption{
Posterior distributions of some parameters for ``$ww_a$Sterile'' model.
Different colors of greenish, grayish, reddish and bluish contours stand for different datasets. 
}
\label{fig:wwa2}
\end{figure}
\begin{figure}
\includegraphics[width=10cm]{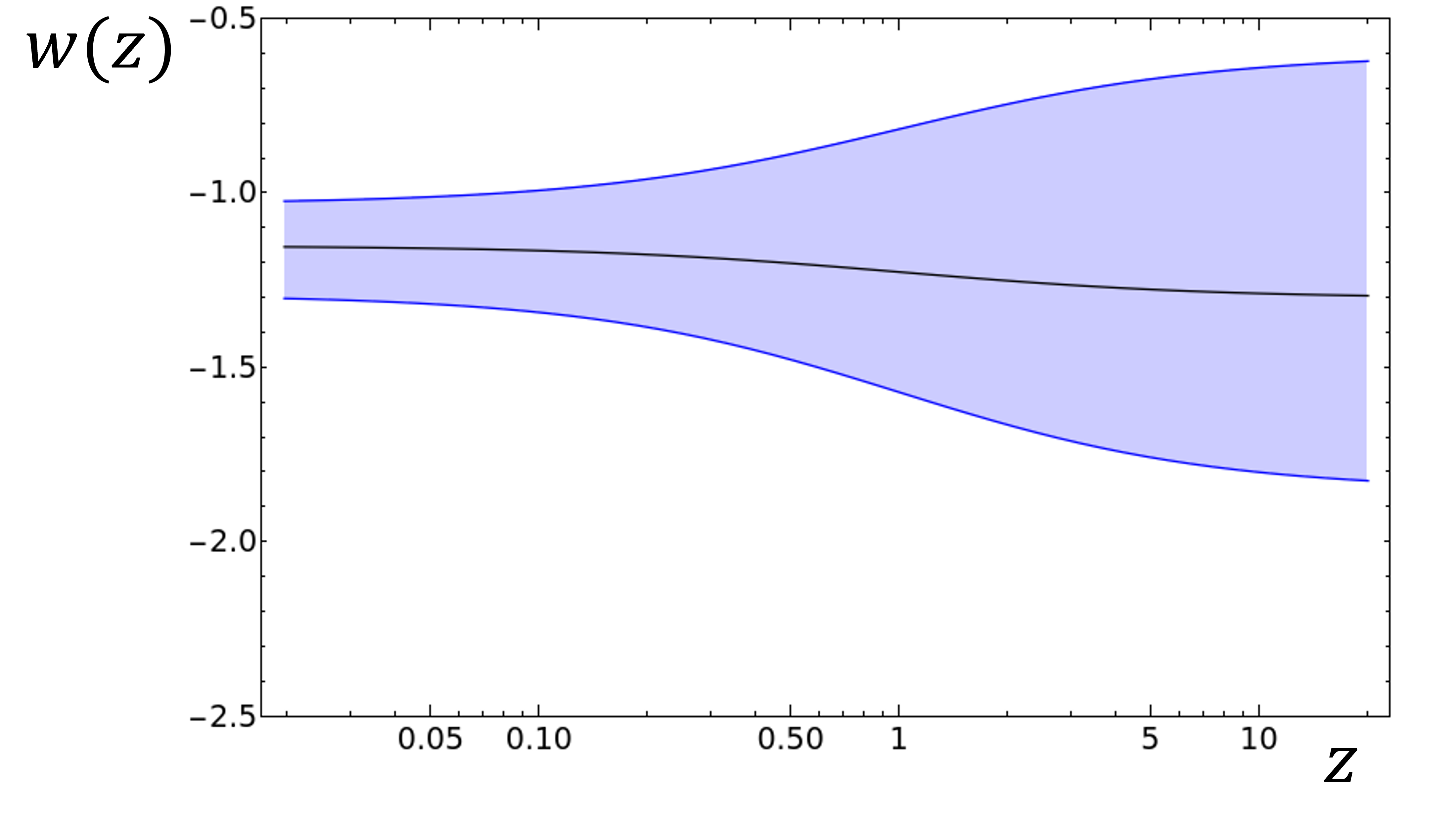}
\caption{The evolution of $w$ with respect to the red-shift $z$ indicated by the dataset {\bf $\mathcal{D}$+R21+DES}. 
The band and the median line correspond to those shown in Table~\ref{tab:Kwwa}.  }
\label{fig:zw}
\end{figure}

\section{Results}
\label{sec:results}

In Tables~\ref{tab:LCDM+Sterile}, \ref{tab:Kwwa}, and \ref{tab:wwa}, we summarize the constraints on the primary and derived parameters of the flat $\Lambda$CDM, $\Lambda$CDM$+$sterile neutrinos (Sterile), $\Omega_K\Lambda$CDM$+$sterile neutrinos ($\Omega_K$Sterile), $w$CDM$+$sterile neutrinos ($w$Sterile) and $w w_a$CDM$+$sterile neutrinos ($w w_a$Sterile) 
for various datasets treating $\mathcal{D}$ ($\equiv$ CMB+BAO) as the base dataset and its combination with Pantheon, DES, R21 prior. While considering  Pantheon, we have not used the R21 prior in order to avoid the twice-counting of a few hundred SNeIa \cite{Dhawan:2020xmp}. 
Triangle plots of the marginalized posterior distributions and two dimensional joint contours of $68\%$ and $95\%$ C.L. are displayed in Fig.~\ref{fig:sterile} (for Sterile); Figs. \ref{fig:OmegaK} and \ref{fig:OmegaK2} (for $\Omega_K$Sterile); Figs. \ref{fig:wsterile} and \ref{fig:wsterile2} (for $w$Sterile); Figs. \ref{fig:wwa}, and \ref{fig:wwa2} (for $w w_a$Sterile). In what follows, we report the key findings of our analyses. Throughout the article, we quantify the level of the tension between two different estimates of the cosmological parameters by using the so-called Gaussian tension, which is defined by
\begin{equation}
T_{x} \equiv \frac{|x_{i}-x_{j}|}{\sqrt{\sigma^2_{x_{i}}+\sigma^2_{x_{j}}}} \,,
\label{tension-estimation}
\end{equation}
where $x_i$ and $x_j$ are the estimated values from two different data $i,j$ with $\sigma_{x_i}$ and $\sigma_{x_u}$ being the 1$\sigma$ errors for each dataset.

\subsection{Sterile}

The $68 \%$ marginalized allowed ranges for the cosmological parameters in $\Lambda$CDM+sterile neutrino model
are summarized in Table.~\ref{tab:LCDM+Sterile}. 
Two dimensional allowed regions and one dimensional posterior distributions are also depicted in Fig.~\ref{fig:sterile}. For comparison, we also show those in the standard $\Lambda$CDM model.
As mentioned above, sterile neutrinos are parameterized by $N_\mathrm{eff}$ and $m_{\nu,\mathrm {sterile}}^{\mathrm{eff}}$.
Interestingly, we obtained a $2\sigma$ lower bound on $m_{\nu,\mathrm{sterile}}^{\mathrm{eff}}$  for the combined dataset $\mathcal{D}$+Pantheon+DES.
This is because as the mass of the sterile neutrino increases, the growth of matter fluctuations with a wave number greater than the free-streaming wave number $k_{\mathrm{fs}}=0.01(m_{\nu_s}/\mathrm{eV})^{1/2}\,\mathrm{Mpc^{-1}}$, is suppressed~\cite{Dodelson:2005tp}, and it would be consistent with DES favoring low $\sigma_8$.
When we analyze the model including R21 prior and DES data  (i.e. for the combined dataset $\mathcal{D}$+R21+DES), the bound on $N_{\rm eff}$ is given as 
$3.2 \lesssim N_\mathrm{eff} \lesssim 3.5$, 
which shows almost the same tendency as the one for the case with $\Lambda$CDM+$N_{\rm eff}$ model reported in the Planck paper~\cite{Planck:2018vyg}\footnote{
If we include the BBN constraint in the prior, the favored range of $N_\mathrm{eff}$ is somewhat decreased~\cite{Seto:2021xua}.
}. 
This indicates that the DES data does not affect the bound on $N_{\rm eff}$ even in the $\Lambda$CDM+sterile neutrino case. When the dataset of  $\mathcal{D}$+Pantheon+DES is considered, the value of $S_8$ is slightly decreased by including sterile neutrinos, which is also observed when usual massive neutrinos is considered \cite{DES:2017myr}.  

\subsection{$\Omega_{K}$Sterile}

The $68 \%$ marginalized allowed ranges for the cosmological parameters in $\Omega_{K}$Sterile model 
are summarized in Table.~\ref{tab:Kwwa}.
Two dimensional allowed regions and one dimensional posterior distributions are also depicted respectively in Figs.~\ref{fig:OmegaK} and \ref{fig:OmegaK2} for the analysis including the R21 prior and the DES data simultaneously and separately.
When we vary $\Omega_K$ in models with sterile neutrino characterized with the parameters $N_\mathrm{eff}$ and $m_{\nu,\mathrm {sterile}}^{\mathrm{eff}}$, a spatially flat Universe $K=0$ is still consistent for datasets without the R21 prior and the DES data.
On the other hand, when we include the R21 prior and DES data,
the favored range of $3.1 \lesssim N_\mathrm{eff} \lesssim 3.3~(1\sigma)$ is obtained. 
Furthermore,  in $\Omega_{K}$Sterile model,  a spatially open Universe is favored at more than $2 \sigma$ and the mean value of $S_8$ coincides with that reported by DES.  

\subsection{$w$Sterile}

The $68 \%$ marginalized allowed ranges for the cosmological parameters of the Sterile+$w$DE model are summarized in Table~\ref{tab:wwa}. Also the two dimensional allowed regions and one dimensional posterior distributions are shown in Figs.~\ref{fig:wsterile} and \ref{fig:wsterile2} respectively for the analysis including the R21 prior and the DES data simultaneously and separately. Indeed, a phantom dark energy ($w < -1$) is allowed for all the datasets. In particular, the evidence of $w < -1$ is strengthened in the presence of the R21 prior, i.e. for both the datasets, namely, {\bf $\mathcal{D}$+R21} and {\bf $\mathcal{D}$+R21+DES}. Moreover, when the DES data is included, a nonzero sterile neutrino mass of the order of sub-eV is found ($m_{\rm \nu, sterile}^{\rm effective} = 0.35^{+0.14}_{-0.23}$ at 68\% CL for {\bf $\mathcal{D}$+Pantheon+DES}). The evidence of a nonzero sterile neutrino mass is also found even for the dataset of $\mathcal{D}$+R21+DES.
Regarding the $H_0$ and $S_8$, we observe that in this extended scenario where $w$ has been considered instead of $\Omega_K$,  the constraints on $S_8$ are almost identical  with those obtained in the $\Omega_K$Sterile scenario, however,  $H_0$ is slightly increased ($\sim$ 2 km/sec/Mpc) in the $w$Sterile case only for the combined datasets $\mathcal{D}$+R21 and $\mathcal{D}$+R21+DES.

\subsection{$ww_a$Sterile}

The $68 \%$ marginalized allowed ranges for the cosmological parameters in $ww_a$Sterile model 
are also summarized in Table~\ref{tab:wwa}. Two dimensional allowed regions and one dimensional posterior distributions are also depicted respectively in Figs.~\ref{fig:wwa} and \ref{fig:wwa2} respectively for the analysis including the R21 prior and the DES data simultaneously and separately.

Even when we take $w$ and $w_a$ as free parameters instead of $\Omega_K$, the resultant $H_0$ and $S_8$ are almost same as those in the $\Omega_K$Sterile model for datasets  without the R21 prior. 
However, when we include the R21 prior and the DES data, a larger $H_0$ and a smaller $S_8$ are inferred, and 
interestingly, a more 
nonzero  sterile neutrino mass of the order of sub-eV is indicated, even compared to that in the  $w$Sterile model. 
This is due to the fact that a time-varying EoS has more degrees of freedom to change the distance to the last scattering surface whose effects are degenerate with those of $N_{\rm eff}$ and sterile neutrino masses.  Finally, in Fig.~\ref{fig:zw}, we show the redshift evolution of the dark energy equation of state $w$, considering the mean values of $w_0$, $w_a$ from the combined analysis  $\mathcal{D}$+R21+DES.  We observe a phantom nature of the dark energy ($w< -1$) in the high redshift, however, $w$ is well consistent with the cosmological constant at the present epoch.

\begin{table}[h]
\begin{tabular}{lccccccc}
\hline 
Parameter & \ensuremath{\Lambda}CDM & $w$CDM & $ww_{a}$CDM & Sterile & \ensuremath{\Omega_{K}}Sterile & \ensuremath{w}Sterile & \ensuremath{ww_{a}}Sterile\tabularnewline
\hline 
{\boldmath\ensuremath{m_{\nu,{\rm {sterile}}}^{{\rm {eff}}}}} & $-$ & $-$ & $-$ & 0.05048 & 0.2068 & 0.017980 & 0.303472\tabularnewline
{\boldmath\ensuremath{N_{\rm eff}}} & $-$ & $-$ & $-$ & 3.1974 & 3.1863 & 3.1822 & 3.1539\tabularnewline
{\boldmath\ensuremath{\Omega_{K}}} & $-$ & $-$ & $-$ & $-$ & 0.006193 & $-$ & $-$\tabularnewline
{\boldmath\ensuremath{w}} & $-$ & $-1.0163$ & $-1.0398$ & $-$ & $-$ & $-1.0232$ & $-1.0636$ \tabularnewline
{\boldmath\ensuremath{w_{a}}} & $-$ & $-$ & $-0.00353$ & $-$ & $-$ & $-$ & $-0.009172$ \tabularnewline
\ensuremath{H_{0}} & 68.90 & 69.32 & 69.77 & 69.30 & 70.16 & 70.05 & 72.5818\tabularnewline
\ensuremath{S_{8}} & 0.7970 & 0.7966 & 0.8035 & 0.7930 & 0.7818 & 0.80052 & 0.776137\tabularnewline
\hline 
\hline 
\ensuremath{\chi_{\mathrm{CMB}}^{2}} & 2771.46 & 2770.08 & 2767.20 & 2773.26 & 2772.22 & 2770.20 & 2770.64\tabularnewline
\ensuremath{\chi_{\mathrm{BAO}}^{2}} & 6.383 & 7.259 & 7.933 & 5.693 & 8.1259 & 7.2362 & 6.975\tabularnewline
\ensuremath{\chi_{\mathrm{H0}}^{2}} & 17.903 & 14.613 & 11.514 & 14.83 & 9.130 & 9.769 & 10.013\tabularnewline
\ensuremath{\chi_{\mathrm{DES}}^{2}} & 507.34 & 507.35 & 507.57 & 507.47 & 505.270 & 507.98 & 504.55\tabularnewline
\ensuremath{\chi_{\mathrm{prior}}^{2}} & 3.1933 & 3.249 & 4.974 & 3.694 & 3.443 & 3.625 & 2.932\tabularnewline
\hline 
\ensuremath{\chi_{\mathrm{Total}}^{2}} & 3306.28 & 3302.55 & 3299.20 & 3304.95 & 3298.19 & 3298.80 & 3295.11\tabularnewline
\ensuremath{\chi_{\mathrm{Total}}^{2}-\chi_{\mathrm{Total \Lambda CDM}}^{2}} & $-$ & $-3.72$ & $-7.08$ & $-1.33$ & $-8.09$ & $-7.48$ & $-11.17$ \tabularnewline
\hline 
\ensuremath{\Delta {\rm AIC}} & $-$
 & $-1.72$ & $-3.08$ & 2.67 & $-2.09$ & $-1.48$ & $-3.17$  \tabularnewline
\hline 
\end{tabular}
\caption{The best-fit values of the cosmological parameters, $\chi^2_{\rm Tot}$ and the $\Delta\mbox{AIC}~[= \mbox{AIC}~({\rm Model}) - \mbox{AIC}~({\Lambda\mbox{CDM}})]$ values for the datasets with R21 and DES (i.e. for the combined dataset $\mathcal{D}$+R21+DES). }
\label{tab:bestfit}
\end{table}

\subsection{Comparison of the models} 

In this section we offer a comparison between various cosmological models with and without sterile neutrinos.  
To compare the goodness of fit in various cosmological models, in Table~\ref{tab:bestfit} we show the best-fit values of some key cosmological parameters, $\chi^2$ for the best fit results and  a model comparison statistics  considering the datasets including the R21 prior and the DES data.

We focus on the best fit values of $H_0$ and $S_8$ obtained in different cosmological scenarios.  When we include the R21 prior and the DES data, we find that the best fit values of $H_0$ in Sterile, $\Omega_K$Sterile and $w$Sterile are quite similar ($H_0 \sim 70$ km/s/Mpc) but the best fit value of $H_0$ in $ww_a$Sterile models attains the maximum ($H_0 \sim 72.6$ km/s/Mpc). We note that the inclusion of the sterile neutrinos affects the expansion history a bit and such effects are  clearly visible when we compare the models $ww_a$CDM and $ww_a$Sterile in the light of the $H_0$ tensions. 
On the other hand, focusing on the $S_8$ parameter, we find that $S_8$ attains the 
lowest best fit value in the $ww_a$Sterile model ($S_8 \sim 0.78$) compared to other models. Thus, we notice that among all the listed cosmological scenarios  with and without the sterile neutrinos (see Table~\ref{tab:bestfit}),  in the $ww_a$Sterile model $H_0$ attains the maximum value and $S_8$ attains the lowest value. In addition, we notice that   $\chi^2_{\rm Tot}$ for $ww_a$Sterile model attains the lowest value.  This shows an improvement in the fit in the $ww_a$Sterile model for this combined dataset with R21 and DES.  

We also perform a model comparison statistic, namely, the Akaike Information criteria (AIC) defined as \cite{Akaike:1974,Liddle:2007fy}: $\mbox{AIC} = \chi^2_{\rm Total} + 2 m$ (where $m$ is the total number of free parameters of the model). At the end of Table~\ref{tab:bestfit} we show the values of $\Delta\mbox{AIC}$ calculated with respect to the reference model $\Lambda$CDM defined as $\Delta\mbox{AIC} = \mbox{AIC}~({\rm Model}) - \mbox{AIC}~({\Lambda\mbox{CDM}})$.  The negative value of $\Delta\mbox{AIC}$ indicates that the model is preferred over the $\Lambda$CDM model.  According to the AIC analysis (see Table~\ref{tab:bestfit}) and the Jeffrey's scale \cite{jeffreys1961theory,Kass:1995loi}, we see that the inclusion of sterile neutrinos does not always offer a better fit compared to the models without sterile neutrinos. For example, although both $w$CDM and $w$Sterile are preferred over $\Lambda$CDM, but in the light of the AIC values, $w$CDM remains in the favored position compared to the $w$Sterile model. However, on the other hand, considering the dynamical DE equation of state, our conclusion alters because according to the results,  $ww_a$Sterile model gives better fit compared to $ww_a$CDM model (i.e. without sterile neutrinos). Although this improvement is very mild, but such improvement is likely due to the inclusion of the sterile neutrinos.  This is not so surprising because an evidence of dynamical DE has been reported by DESI 2024 \cite{DESI:2024kob}. Therefore, we noticed that for this particular dataset, $ww_a$CDM model (with and without sterile neutrinos)  state seems to be a very appealing candidate 
for  further investigations. In fact, models with varying dark energy equation of state with or without the sterile neutrinos may offer new avenues in the light of the cosmological tensions.

To quantify how the tension is reduced in these models, we estimate and display the Gaussian tension for $H_0$ and $S_8$ in Table~\ref{tab:sigma} for the dataset $\mathcal{D}$+Pantheon, that means without considering the R21 and DES data. 
The Gaussian tensions for $H_0$ and $S_8$ are respectively calculated as 

\begin{equation}  \sigma_{H0}=\frac{H_{0~\mathrm{Planck\&BAO\&Pantheon}}-73.
30}{\sqrt{\sigma_{\mathrm{Planck\&BAO\&Pantheon}}^{2}+1.04^{2}}} \,,
\end{equation}
for the Hubble tension with direct measurement~\cite{Riess:2021jrx} and
\begin{equation}
\sigma_{S8}=\frac{S_{8~\mathrm{Planck\&BAO\&Pantheon}}-0.776}{\sqrt{\sigma_{\mathrm{Planck\&BAO\&Pantheon}}^{2}+\frac{0.017^{2}+0.017^{2}}{2}}} \,,
\end{equation} 
for the $S_{8}$ tension with DES~\cite{DES:2021wwk}. 
We find that in the $\Omega_K$Sterile model,  and $ww_a$Sterile model both $H_{0}$ tension and $S_{8}$ tension are reduced with
a significance of over 1$\sigma$ from $\Lambda$CDM model. In the $w$Sterile model, the $H_0$ tension is also reduced with a significance of over 1$\sigma$.~\footnote{We quote the 68\% confidence limits $H_0 = 68.31\pm0.82$ and $S_8 = 0.829\pm0.011$ for $ww_a$ model by $\mathcal{D}$+Pantheon~\cite{Planck:2018vyg}. Therefore, we find that the relieve of the Hubble tension is mostly due to the introduction of $w_a$DE, while the relieve of the $S_8$ tension is due to the introduction of the sterile neutrino.}
We again stress here that, in the $ww_a$Sterile model, the mass of sterile neutrino in sub-eV scale is preferred.

\begin{table}
\begin{tabular}{lccccccc}
\hline 
 Parameter  &  \ensuremath{\Lambda}CDM  & $w$CDM & $ww_a$CDM&  Sterile  &  \ensuremath{\Omega_{K}} Sterile  & \ensuremath{w}Sterile &  \ensuremath{ww_{a}}Sterile \\
\hline  \ensuremath{H_{0}}  &  \ensuremath{67.74\pm0.42}  & $68.26\pm 0.82$ & $68.32\pm 0.82$&\ensuremath{67.89_{-0.69}^{+0.40}}  &  \ensuremath{68.25_{-0.80}^{+0.68}}  & \ensuremath{68.75\pm0.92} &  \ensuremath{68.76_{-0.93}^{+0.84}} \\
 \ensuremath{\sigma_{H0}}  &  4.95\ensuremath{\sigma\ }  & 3.81\ensuremath{\sigma\ }& 3.76\ensuremath{\sigma\ }& 4.57\ensuremath{\sigma\ }  &  3.95\ensuremath{\sigma}  &  3.28\ensuremath{\sigma}  &  3.32\ensuremath{\sigma\ } \\
\hline  \ensuremath{S_{8}}  &  \ensuremath{0.8234_{-0.011}^{+0.0098}}  & \ensuremath{0.827\pm 0.013} & \ensuremath{0.833\pm 0.014}&\ensuremath{0.812_{-0.014}^{+0.019}}  &  \ensuremath{0.809_{-0.014}^{+0.021}}  & 0.81\ensuremath{0_{-0.013}^{+0.021}} &  \ensuremath{0.809_{-0.016}^{+0.022}} \\
 \ensuremath{\sigma_{S8}}  &  2.38\ensuremath{\sigma\ } &  2.38\ensuremath{\sigma\ } & 2.59\ensuremath{\sigma\ } &  1.51\ensuremath{\sigma\ }  &  1.34\ensuremath{\sigma}  & 1.40\ensuremath{\sigma}  &  1.29\ensuremath{\sigma\ } \\
 \hline 
\end{tabular}
\caption{$68\%$ limits of $H_{0}$ and $S_{8}$, and the Gaussian tension
for the $H_{0}$ tension and $S_{8}$ tension. To obtain this table, we use the dataset $\mathcal{D}+$Pantheon.}\label{tab:sigma} 
\end{table}

\section{Summary and Conclusions}
\label{sec:summary}

In this paper, we have investigated constraints on the sterile neutrinos in the context of various 
cosmological scenarios in the light of the cosmological tensions such as the $H_0$ and $S_8$ tensions. 
When we do not include the data from R21 and DES, we only obtain an upper bound on $m^\mathrm{eff}_{\nu, {\rm sterile}}$ and $N_\mathrm{eff}$ is close to $N_\mathrm{eff}^\mathrm{SM}$ 
within $2\sigma$ in all models we considered in this paper, i.e., $\Lambda$CDM, Sterile, $\Omega_K$Sterile, $w$Sterile and $ww_a$Sterile models. It should be noted that, even in the presence of the sterile neutrinos, the spatial curvature in the $\Omega_K$Sterile model is consistent with flat Universe within $1\sigma$, the dark energy EoS is also consistent with $w=-1$.

However, when the DES data is included in the analysis, nonzero sterile neutrino masses is favored at 1$\sigma$ level in Sterile, 
$\Omega_K$Sterile, $w$Sterile models, and at 2$\sigma$ in $ww_a$Sterile model. When the DES data and R21 are both included, we only obtain an upper bound on $m^\mathrm{eff}_{\nu, {\rm sterile}}$ at 2$\sigma$ for Sterile, 
$\Omega_K$Sterile, $w$Sterile models, while nonzero sterile neutrino masses are still favored at 2$\sigma$ level in  $ww_a$Sterile model.  When R21 is included, the value of $H_0$ tends to be increased compared to that in the $\Lambda$CDM case in all extended models, and when the DES is included, $S_8$ decreases compared to the one without DES, and hence, in the $ww_a$Sterile model, nonzero sterile neutrino masses are inferred with the cosmological tensions being reduced to some extent. 

It should also be mentioned that, in the presence of sterile neutrinos, negative spatial curvature of the Universe is favored at 2$\sigma$ level when R21 and the DES data are included\footnote{
One can indeed construct an inflationary model with $\Omega_K>0$ ~\cite{Linde:1998iw,Linde:1999wv}.
}. As we argued in this paper, the cosmological tensions can be reduced once we assume the existence of sterile neutrino in some extended frameworks such as nonflat Universe and a varying dark energy EoS model with nonzero sterile neutrino masses.  This implies that the current cosmological data do not exclude the existence of sterile neutrinos, or even suggest its existence in some extended models.

\section*{Acknowledgments}
The authors thank the referee for some important comments. 
SP acknowledges the financial support from  the Department of Science and Technology (DST), Govt. of India, under the Scheme
``Fund for Improvement of S\&T Infrastructure (FIST)'' [File No. SR/FST/MS-I/2019/41].
This work was supported by JSPS KAKENHI Grant No. 19K03860, No. 19K03865, No. 23K03402 (OS), No. 19K03874, No. 23K17691 (TT), MEXT KAKENHI No. 23H04515 (TT), and JST SPRING, Grant No. JPMJSP2119 (YT).

\bibliography{ref}

\end{document}